\documentstyle[11pt]{article}

\newlength{\minitwocolumn}
\catcode`\@=11
\def\relaxnext@{\let\next\relax}
\font\tenmsy=msym10 scaled\magstep1
\font\sevenmsy=msym7 scaled\magstep1
\font\fivemsy=msym5  scaled\magstep1
\newfam\msyfam
\textfont\msyfam=\tenmsy
\scriptfont\msyfam=\sevenmsy
\scriptscriptfont\msyfam=\fivemsy
\font\teneuf=eufm10 scaled\magstep1
\font\seveneuf=eufm7 scaled\magstep1
\font\fiveeuf=eufm5 scaled\magstep1
\newfam\euffam
\textfont\euffam=\teneuf
\scriptfont\euffam=\seveneuf
\scriptscriptfont\euffam=\fiveeuf
\def\frak{\relaxnext@\ifmmode\let\next\frak@\else
 \def\next{\Err@{Use \string\frak\space only in math mode}}\fi\next}
\def\goth{\relaxnext@\ifmmode\let\next\frak@\else
 \def\next{\Err@{Use \string\goth\space only in math mode}}\fi\next}
\def\frak@#1{{\frak@@{#1}}}
\def\frak@@#1{\noaccents@\fam\euffam#1}
\def\Bbb{\relaxnext@\ifmmode\let\next\Bbb@\else
 \def\next{\Err@{Use \string\Bbb\space only in math mode}}\fi\next}
\def\Bbb@#1{{\Bbb@@{#1}}}
\def\Bbb@@#1{\noaccents@\fam\msyfam#1}
\def\accentfam@{7}
\def\noaccents@{\def\accentfam@{0}}
\catcode`\@=\active
\newcommand{\bz}{{\Bbb Z}}

\newcommand{\bc}{{\Bbb C}}

\makeatletter
\@addtoreset{equation}{section}
\makeatother

\newtheorem{thm}{Theorem}[section]
\newtheorem{prop}[thm]{Proposition}
\newtheorem{lem}[thm]{Lemma}

\begin{document}
\begin{flushright}
RIMS-972 \\
April 1994
\end{flushright}
\vspace{40pt}
\begin{center}
\begin{Large}
{\bf Quantum Knizhnik--Zamolodchikov equation \\
for $U_q \bigl(\widehat{\frak s \frak l _{n}}\bigr)$ and
integral formula}
\end{Large}
\vspace{12pt}

\vspace{72pt}
 Quantum Knizhnik--Zamolodchikov equation \\
for $U_q \bigl(\widehat{\frak s \frak l _{n}}\bigr)$ and
integral formulaTakeo Kojima\raisebox{2mm}{\small 1} and Yas-Hiro Quano\raisebox{2mm}
{\small 1,2}

\vspace{6pt}

$~^1$ {\it Research Institute for Mathematical Sciences,
     Kyoto University, Kyoto 606, Japan}

$~^2$ {\it Institute for Nuclear Study, University of Tokyo,
     Tanashi, Tokyo 188, Japan}
\vspace{72pt}

\underline{Abstract}

\end{center}

Presented is an integral formula for solutions
to the quantum Knizhnik--Zamolodchikov equation
of level $0$ associated with
the vector representation of
$U_q \bigl(\widehat{\frak s \frak l _{n}}\bigr)$.
This formula gives a generalization of both
our previous work for
$U_q \bigl(\widehat{\frak s \frak l _{2}}\bigr)$
and Smirnov's formula for form factors
of $SU(n)$ chiral Gross-Neveu model.

\vspace{60pt}

\newpage
\section{Introduction}

In our previous paper \cite{JKMQ}
we gave an integral formula for solutions to
the quantum Knizhnik--Zamolodchikov (KZ) equation \cite{FR}
for the quantum affine algebra
$U_q\bigl({\widehat{\frak s \frak l}}_{2}\bigr)$
when the spin is $1/2$,
the level is $0$ and $|q|<1$.
The present paper is a
$U_q\bigl({\widehat{\frak s \frak l}}_{n}\bigr)$
generalization of \cite{JKMQ}.
Our idea is based on Ref.\cite{Smbk}.
Instead of solving the quantum KZ equation
we consider a system of difference equations 
for a vector-valued function
in $N$ variables $(z_1,\cdots,z_N)$
which takes values in the $N$-fold 
tensor product of the vector
representation $V={\bc}^n$ of
$U_q\bigl({\widehat{\frak s \frak l}}_{n}\bigr)$.

For a fixed complex number $q$ satisfying $0<|q|<1$,
let $R(z)\in {\rm End}(V\otimes V)$ be
the standard trigonometric $R$-matrix
associated with the vector representation
$V\cong \bc v_1 \oplus \cdots \oplus \bc v_n $
of $U_q\bigl({\widehat{\frak s \frak l}}_{n}\bigr)$.
The matrix $R(z)$
satisfying the Yang-Baxter equation
and the unitarity relation
is given as follows:
$$
R(z)v_{\varepsilon'_1}\otimes v_{\varepsilon'_2}=
\sum_{\varepsilon_1,\varepsilon_2} 
v_{\varepsilon_1}\otimes v_{\varepsilon_2}
R^{\varepsilon_1\varepsilon_2}_{
   \varepsilon'_1\varepsilon'_2}(z),
$$
where the nonzero entries are
$$
\begin{array}{rcl}
R^{\varepsilon \varepsilon}_{
   \varepsilon \varepsilon} (z) & = & 1, \\
R^{\varepsilon \varepsilon'}_{
   \varepsilon \varepsilon'} (z) & = & 
b(z)=\displaystyle{(1-z)q \over 1-zq^2},
\quad \mbox{if $\varepsilon \neq \varepsilon'$,} \\
R^{\varepsilon \varepsilon'}_{
   \varepsilon' \varepsilon} (z) & = &
\left\{ \begin{array}{ll}
\displaystyle{(1-q^2)z \over 1-zq^2 } & 
\mbox{if $\varepsilon < \varepsilon'$,} \\
\displaystyle{(1-q^2) \over 1-zq^2 } & 
\mbox{if $\varepsilon > \varepsilon'$.}
\end{array} \right.
\end{array}
$$
In statistical mechanics language
each entry of the $R$-matrix is a local Boltzmann weight
for a single vertex with bond states
$i, j, k, l \in \bz _n$:

\begin{minipage}{\minitwocolumn}
\unitlength 1mm
\begin{picture}(40,30)
\put(47,15){$R^{ik}_{jl}(z_1 /z_2 )=
$}
\end{picture}
\end{minipage}%
\hspace{\columnsep}%
\begin{minipage}{\minitwocolumn}
\unitlength 1mm
\begin{picture}(40,30)
\put(-5,6){\begin{picture}(101,0)
\put(0,10){\vector(1,0){15}}
\put(15,10){\line(1,0){5}}
\put(10,0){\vector(0,1){15}}
\put(10,15){\line(0,1){5}}
\put(-3,9.5){$j$}
\put(9,21){$k$}
\put(9,-4){$l$}
\put(21,9.5){$i$}
\put(13,6){$z_{1}$}
\put(11,14){$z_{2}$}
\put(24,9){$,$}
\end{picture}
}
\end{picture}
\end{minipage}

\noindent where each line carries a spectral parameter.

In what follows we shall work with
the tensor product of finitely many $V$'s.
Following the usual convention
we let $R_{jk}(z)$ ($j\neq k$)
signify the operator on
$V^{\otimes N}$ acting as $R(z)$
on the $(j,k)$-th tensor
components and as identity
on the other components.
In particular we have
$R_{kj}(z)=P_{jk}R_{jk}(z)P_{jk}$, where
$P\in {\rm End}(V\otimes V)$
stands for the transposition
$P(x\otimes y)=y\otimes x$.

The equations we are concerned with
in this paper are
(1) $R$-matrix symmetry and
(2) deformed cyclicity, for a function
$G(z_1,\cdots,z_N) \in V^{\otimes  N}$:
\begin{eqnarray}
(1) && P_{j\,j+1} G (\cdots,z_{j+1},z_j,\cdots)
\quad =
R_{j\,j+1}(z_j/z_{j+1})G (\cdots,z_j,z_{j+1},\cdots),
\label{eqn:R-symm} \\
(2) && P_{12}\cdots P_{N-1 N} G (z_2,\cdots,z_N, z_1 q^{-2n})
=
D_1 G (z_1,\cdots,z_N).
\label{eqn:cyc}
\end{eqnarray}
In (\ref{eqn:cyc}) $D_1$ is an operator
acting on the first component as
$D={\rm diag}(\delta _1 , \cdots, \delta _n )$,
whose entries will be specified below,
and as identity on the other ones.
These are two of the axioms that form factors
in integrable models should satisfy \cite{Sm1}.
Smirnov \cite{Sm1} also pointed out that (\ref{eqn:R-symm})
and (\ref{eqn:cyc}) imply the quantum
KZ equation of level $0$ \cite{FR}
\begin{equation}
\begin{array}{l}
G (z_1,\cdots,z_jq^{2n},\cdots,z_N)
=
R_{j-1\,j}(z_{j-1}/z_j q^{2n} )^{-1}
\cdots
R_{1\,j}(z_{1}/z_j q^{2n})^{-1}D_j^{-1} \\
\qquad\qquad \times
R_{j\,N}(z_{j}/z_N)
\cdots
R_{j\,j+1}(z_{j}/z_{j+1})
G (z_1,\cdots,z_j,\cdots,z_N).
\label{qKZ}
\end{array}
\end{equation}
Throughout this article the functions we consider
are not necessarily single valued in
$z_j$ but are meromorphic in the variable
$\log z_j$.
Accordingly the shift
$z_j\rightarrow z_jq^{-2n}$
as in (\ref{eqn:cyc}) is understood to mean
$\log z_j\rightarrow \log z_j-2n\log q$.

In the sequel we set $\tau=q^{-1}$.
Define the components of $G$ by
\begin{equation}
G(z _1 , \cdots , z _N ) =
\sum_{\varepsilon _j =1}^{n}
v_{\varepsilon _1 } \otimes \cdots
\otimes v_{\varepsilon _N }
G^{\varepsilon _1 \cdots \varepsilon _N }
(z _1 , \cdots , z_N ). \label{eqn:components}
\end{equation}
Then the equation (\ref{eqn:R-symm}) reads as
\begin{eqnarray}
G ^{\cdots \varepsilon \varepsilon \cdots}
(\cdots,z_j,z_{j+1},\cdots)
&=&
G ^{\cdots \varepsilon \varepsilon \cdots}
(\cdots,z_{j+1},z_j,\cdots),
\label{eqn:Rsym1} \\
G ^{\cdots \varepsilon \varepsilon' \cdots}
(\cdots,z_j,z_{j+1},\cdots)
&=&
{z_j-z_{j+1}\tau^2 \over (z_j-z_{j+1})\tau }
G ^{\cdots \varepsilon' \varepsilon \cdots}
(\cdots,z_{j+1},z_j,\cdots) \nonumber \\
&-&
{(1-\tau^2)z_{j j+1}^{\varepsilon \varepsilon'} 
                      \over (z_j-z_{j+1})\tau}
G^{\cdots \varepsilon' \varepsilon \cdots}
(\cdots,z_j,z_{j+1},\cdots),
\label{eqn:Rsym}
\end{eqnarray}
where
$$
z_{j j+1}^{\varepsilon \varepsilon'}=
\left\{ \begin{array}{ll}
z_j & \mbox{if $\varepsilon <\varepsilon'$,} \\
z_{j+1} & \mbox{if $\varepsilon > \varepsilon'$,}
\end{array} \right.
$$
and (\ref{eqn:cyc}) reads as
\begin{equation}
G ^{\varepsilon_2 \cdots \varepsilon_N \varepsilon_1} 
(z_2,\cdots,z_N,z_1 \tau^{2n})
=
\delta_{\varepsilon_1}~
G ^{\varepsilon_1 \varepsilon_2 
    \cdots \varepsilon_N} (z_1,z_2,\cdots,z_N).
\label{eqn:cyclic}
\end{equation}
Note that the singularity at $z_j =z_{j+1}$
in (\ref{eqn:Rsym}) is spurious.
The equations (\ref{eqn:Rsym1}--\ref{eqn:cyclic})
split into blocks, each involving components such that
$$
\sharp \{j| \varepsilon_j =i\}=m_i, \quad N=\sum_{i=1}^{n} m_i .
$$
In the present paper we restrict ourselves to
the case $m_1 =\cdots  =m_n =m$ and hence $N=mn$.
According to this restriction, we set 
$\delta_i =\tau^{m(1-n)+2(1-i)}$. 

We use the abbreviation $z^{(j)}=(z^{(j)}_1,\cdots,z^{(j)}_m)$.
Consider the extreme component
\begin{equation}
G^{\overbrace{n\cdots n}^m \cdots \overbrace{1\cdots 1}^m}
(z^{(n)}|\cdots |z^{(1)}_{1})
=H(z^{(n)}|\cdots |\,z^{(1)}).
\label{eqn:1_comp}
\end{equation}
Because of (\ref{eqn:Rsym1})
this function is symmetric
in the variables
$z^{(1)}$'s, $\cdots $, $z^{(n)}$'s, separately. 
The equation (\ref{eqn:Rsym}) tells that
all the components with fixed $m$
are uniquely determined from $H$.
Conversely given any such $H$
the Yang-Baxter equation guarantees
that (\ref{eqn:R-symm}) can be
solved consistently
under the condition (\ref{eqn:1_comp}).

We wish to find an integral formula of the form
\begin{equation}
H(z_1 , \cdots , z_N )=
(S_{M N} F)(z_1,\cdots ,z_N),
\label{DefH}
\end{equation}
where $S_{M N}$ stands for the following integral transform
\begin{equation}
(S_{M N} F)(z_1,\cdots ,z_N)=
\oint_{C} dx_1 \cdots \oint_{C} dx_M
F(x_1 , \cdots , x_M | z_1 , \cdots , z_N )
\Psi (x_1 , \cdots , x_M | z_1 , \cdots , z_N ).
\label{intform}
\end{equation}
The notation is explained below.

The kernel
$\Psi$
has the form
\begin{equation}
\Psi (x_1 , \cdots , x_M | z_1 , \cdots , z_N )=
\vartheta (x_1 , \cdots , x_M | z_1 , \cdots , z_N )
\prod_{\mu =1}^{M}
\prod_{j=1}^{N} \psi \Bigl(\frac{x_{\mu }}{z_j }\Bigr),
\end{equation}
where
\begin{equation}
\psi(z)=\frac{1}{(zq^{n-1};q^{2n})_{\infty}
(z^{-1}q^{n-1};q^{2n})_{\infty}},
\qquad
(z;p)_\infty=\prod_{k=0}^\infty(1-z p^k).
\end{equation}
Assume that
the function $\vartheta$
is anti-symmetric and holomorhpic
in the $x_\mu \in \bc \backslash \{0\}$,
and is symmetric and meromorphic
in the $\log z_j\in \bc$,
possessing the following transformation property
\begin{equation}
\begin{array}{rcl}
\vartheta (x_1 , \cdots , x_M |
z_1 , \cdots , z_j \tau^{2n} ,\cdots,z_N )
& = &
\vartheta (x_1 , \cdots , x_M | z_1 , \cdots , z_N )
\displaystyle\prod_{\mu =1}^{M}
\frac{-z_j \tau^{n-1}}{x_{\mu }}, \\
\vartheta (x_1 , \cdots, x_{\mu}\tau^{2n} ,
\cdots, x_M | z_1 , \cdots , z_N )
& = &
\vartheta (x_1 , \cdots , x_M | z_1 , \cdots , z_N )
\displaystyle\prod_{j=1}^{N} 
\frac{-x_{\mu } \tau^{n-1}}{z_j }.
\end{array}
\label{pstr}
\end{equation}
The function
$\vartheta$
is otherwise arbitrary,
and the choice of $\vartheta $'s corresponds to that of
solutions.
The integration $\oint_{C} dx_{\mu}$
is along a simple closed curve $C=C(z_1 ,\cdots , z_N)$ 
oriented anti-clockwise,
which encircles the points
$z_j \tau ^{-n+1-2nk} (1\leq j \leq N , k\geq 0)$
but not
$z_j \tau ^{n-1+2nk} (1\leq j \leq N , k\geq 0)$.
Finally
\begin{equation}
F(x_1 , \cdots , x_M | z^{(n)}|\cdots |z^{(1)})
=\frac{\Delta ^{(m)}
(x_1 , \cdots , x_M | z^{(n)}|\cdots |z^{(1)})}
{\displaystyle\prod_{k,k'=1 \atop k<k'}^{n} 
 \prod_{j=1}^{m}\prod_{j'=1}^{m}
       (z^{(k)}_{j}-z^{(k')}_{j'} \tau ^2 )},
\label{def:F}
\end{equation}
where $\Delta^{(m)}$ is
a certain homogeneous polynomial to be determined,
antisymmetric in the variables $(x_1 ,\cdots , x_n )$ and
symmetric in the variables $z^{(1)}$'s, $\cdots$,
$z^{(n)}$'s, separately.

The rest of the paper is organized as follows.
In Section 2 we introduce a special basis of $V$
in terms of the quantum monodromy operators.
In Section 3 we describe the main theorem 
of the present paper.
The subsequent two sections are devoted to proof of it.
Section 4 is for $m=1$ and Section 5 is for general case.
In section 6 we discuss the relation 
among other works and ours.

~

\section{Quantum monodromy operators and the special basis}

We shall construct a special basis
$\{ w_{\alpha_1 \cdots \alpha_N}
(z_1,\cdots,z_N)\}$ with $
\alpha_j \in \{1,2,\cdots,n\}$
of $V^{\otimes N}$ depending on the parameter $(z_1,\cdots,z_N)$,
which satisfy
\begin{equation}
P_{j j+1}w_{\cdots \alpha_{j+1} \alpha_j \cdots} 
(\cdots,z_{j+1},z_j,\cdots)
=R_{j j+1}(z_j/z_{j+1})
w_{\cdots \alpha_j \alpha_{j+1} \cdots}
(\cdots,z_j,z_{j+1},\cdots).
\label{R-goog}
\end{equation}
The procedure goes as follows.
Define the quantum monodromy operator
${\cal T}_{\varepsilon \varepsilon'}
(z_1,\cdots,z_N| t) 
\in {\rm End} (V^{\otimes N})$  by
\begin{equation}
R_{1\,N+1}(z_1/t)\cdots R_{N\,N+1}(z_N/t)
=
\left({\cal T}_{\varepsilon \varepsilon'} 
(z_1,\cdots,z_N| t)\right)_{ 
1\leq \varepsilon, \varepsilon'\leq n}.
\label{eqn:defT}
\end{equation}
Here the $n\times n$ matrix structure is
defined relative to the base
$v_1, \cdots , v_n$ of
the $(N+1)$-th tensor component
of $V^{\otimes (N+1)}$:

\begin{minipage}{\minitwocolumn}
$$
\hfill
{\cal T}_{\varepsilon \varepsilon'} 
(z_1,\cdots,z_N| t)=
$$
\end{minipage}%
\hspace{\columnsep}%
\begin{minipage}{\minitwocolumn}
\unitlength 1mm
\begin{picture}(50,40)
\put(-25,-2){\begin{picture}(101,0)
\put(60,20){\vector(-1,0){60}}
\put(5,21){$\varepsilon$}
\put(55,21){$\varepsilon'$}
\put(15,10){\vector(0,1){20}}
\put(16,24){$z_{1}$}
\put(25,10){\vector(0,1){20}}
\put(26,24){$z_{2}$}
\put(45,10){\vector(0,1){20}}
\put(46,24){$z_{N}$}
\put(33,24){$\cdots$}
\put(35,16){$t$}
\end{picture}
}
\end{picture}
\hfill
\end{minipage}

For $\alpha=(\alpha_1,\cdots,\alpha_N)$
with $\alpha_j\in \{1, \cdots , n\}$, set
$J^{\alpha}_i=\{~j~|~\alpha_j=i\}$ and
$\{ z_1^{(i)}, \cdots , z_m ^{(i)} \}= \{ z_j | \alpha_j =i \}$.
where $1\leq i \leq n$,
and set 
\begin{equation}
\begin{array}{rcl}
w_\alpha(z_1,\cdots,z_N)&=&\displaystyle\prod_{l=1}^{m}~
{\cal T}_{1 n}(z_1,\cdots,z_N |z_{l}^{(n)})
\cdots
\prod_{l=1}^{m}~{\cal T}_{1 2}(z_1,\cdots,z_N |z_{l}^{(2)})
\Omega,\\
\Omega&=&\displaystyle v_1\otimes \cdots \otimes v_1
\quad \in V^{\otimes N}.
\end{array}
\label{eqn:defw}
\end{equation}
Then
$w_\alpha(z_1,\cdots,z_N)$ is visualized as follows:

\begin{minipage}{\minitwocolumn}
$$
\hspace{-5mm}
w_\alpha(z_1,\cdots,z_N)=\sum v_{i_1} \otimes \cdots \otimes v_{i_N}
$$
\end{minipage}%
\hspace{\columnsep}%
\begin{minipage}{\minitwocolumn}
\unitlength 1mm
\begin{picture}(50,100)
\put(-7,10){\begin{picture}(101,0)
\put(70,10){\vector(-1,0){70}}
\put(35,5){$z^{(2)}_{m}$}
\put(70,25){\vector(-1,0){70}}
\put(35,20){$z_1^{(2)}$}
\put(35,13){$\ddots$}
\put(70,55){\vector(-1,0){70}}
\put(35,50){$z_m^{(n)}$}
\put(70,70){\vector(-1,0){70}}
\put(35,65){$z_1^{(n)}$}
\put(35,58){$\ddots$}
\put(5,11){$1$}
\put(65,11){$2$}
\put(5,26){$1$}
\put(65,26){$2$}
\put(5,56){$1$}
\put(65,56){$n$}
\put(5,71){$1$}
\put(65,71){$n$}
\put(2,40){\vdots}
\put(68,40){\vdots}
\put(10,0){\vector(0,1){80}}
\put(20,0){\vector(0,1){80}}
\put(60,0){\vector(0,1){80}}
\put(9,-4){$1$}
\put(9,82){$i_1$}
\put(11,40){$z_{1}$}
\put(19,-4){$1$}
\put(19,82){$i_2$}
\put(21,40){$z_{2}$}
\put(59,-4){$1$}
\put(59,82){$i_N$}
\put(61,40){$z_{N}$}
\put(35,38){$\ddots$}
\put(35,-4){$\cdots$}
\put(35,82){$\cdots$}
\put(72,40){$.$}
\end{picture}
}
\end{picture}
\end{minipage}

Let us introduce the ordered indices set
$(\alpha_1,\cdots,\alpha_N) > 
(\varepsilon_1,\cdots,\varepsilon_N)$
if and only if
$\alpha_i=\varepsilon_i, 
(1 \leq i \leq k)$ and $\alpha_{k+1}>\varepsilon_{k+1}$.
Define the components of the basis by
\begin{equation}
w_{\alpha_1 \cdots \alpha_N}(z_1,\cdots,z_N)
=\sum_{\varepsilon_1,\cdots,\varepsilon_N}
v_{\varepsilon_1}\otimes \cdots \otimes v_{\varepsilon_N}
w_{\alpha_1 \cdots \alpha_N}^{ 
   \varepsilon_1 \cdots \varepsilon_N}
(z_1,\cdots,z_N),
\end{equation}
then we have
\begin{equation}
w_{\alpha_1 \cdots \alpha_N}^{ 
   \varepsilon_1 \cdots \varepsilon_N}
(z_1,\cdots,z_N)=0, \mbox{~~~~if }~~
(\alpha_1,\cdots,\alpha_N) < 
(\varepsilon_1,\cdots,\varepsilon_N).
\label{triang}
\end{equation}
Furthermore for
$\beta=(n,\cdots,n,\cdots,2,\cdots,2,1,\cdots,1)$
we have
\begin{equation}
w^{\beta}_{\beta}(z_1 , \cdots , z_N )
=\prod_{k\in J_i^{\beta},l\in J_j^{\beta} 
        \atop i<j}b(z_k/z_l).
\label{bound}
\end{equation}
{}From (\ref{R-goog}), (\ref{triang}) and (\ref{bound})
we obtain the following explicit formula

\[
G(z_1,\cdots,z_N)=\sum_\alpha w_\alpha(z_1,\cdots,z_N)
H\bigl(\{z_j\}_{j\in J^{\alpha}_n}\,|\cdots |\,\{z_j\}_{j\in
J^{\alpha}_1}\bigr)
\prod_{\alpha_i < \alpha_j}
{1 \over b(z_i/z_j)}.
\]

For $2\leq p \leq n$ and $n\geq i_1 > \cdots > i_p \geq 1$
set
$$
v^{(i_1 \cdots i_p )}=
\sum_{\sigma \in \frak S _p } (-\tau)^{l(\sigma)}
v_{\sigma (i_1)} \otimes \cdots \otimes v_{\sigma (i_p)},
$$
where $l(\sigma)$ is the minimum 
number of permutations such that
$$
v_{\sigma (i_1)} \otimes \cdots \otimes v_{\sigma (i_p)}
=\left( \prod P_{j j+1} \right) v_{i_1} 
\otimes \cdots \otimes v_{i_p}.
$$
For example
$$
v^{(21)}=v_2 \otimes v_1 - \tau v_1 \otimes v_2 ,
$$
$$
\begin{array}{rcl}
v^{(321)}&=&v_3 \otimes v_2 \otimes v_1
-\tau (v_3 \otimes v_1 \otimes v_2 + 
v_2 \otimes v_3 \otimes v_1 )\\
&+&\tau^2 (v_1 \otimes v_3 \otimes v_2 +
v_2 \otimes v_1 \otimes v_3 )
-\tau^3 v_1 \otimes v_2 \otimes v_3 .
\end{array}
$$
For $i\in \{ i_1 , \cdots , i_p \}$
one can easily check the following formulae
\begin{equation}
R_{1 p+1}(z) R_{2 p+1}(z\tau^2 )\cdots 
R_{p p+1}(z\tau^{2p-2})
v^{(i_1 \cdots i_p )} \otimes v_i
=\prod_{j=1}^{p-1} b(z\tau^{2j}) 
v^{(i_1 \cdots i_p )}\otimes v_i .
\label{Chq}
\end{equation}

The poles of $w_{\alpha}(z)$ exist only at
$z_j = z_i \tau^2$ for $i<j$ and $\alpha_i > \alpha_j$.
Thus we have
\begin{equation}
\begin{array}{cl}
&{\rm Res}_{z_{2}=z_{1}\tau^2}\cdots 
{\rm Res}_{z_{n}=z_{n-1}\tau^2}
w_{\alpha_1 \cdots \alpha_n}(z_1,\cdots,z_{n}) \\
=&
\delta_{\alpha_1 n} \delta_{\alpha_2 n-1} \cdots \delta_{\alpha_n 1}
z_1^{n-1} (\tau^2 -1)^{n-1} \tau^{(n-1)^2}
\left[ n-1 \right] ! ~~ v^{(n\cdots 21)},
\end{array}
\end{equation}
where
$$
\left[ k \right] ! = 
\left[ k \right] \cdots \left[ 1 \right], ~~~~
\left[ k \right] = 
\frac{\tau^k -\tau^{-k}}{\tau -\tau^{-1}}.
$$
Furthermore we obtain the recursive residue formula
\begin{equation}
\begin{array}{cl}
&{\rm Res}_{z_{2}=z_{1}\tau^2} \cdots
  {\rm Res}_{z_{n}=z_{n-1}\tau^2}
   w_{\alpha_1 \cdots \alpha_N }(z_1,\cdots,z_{N}) \label{last} 
=\prod_{\alpha_i>\alpha_j \atop i\leq n<j}
b(z_{j}/z_i) \\ 
\times & 
\left( {\rm Res}_{z_{2}=z_{1}\tau^2} 
\cdots {\rm Res}_{z_{n}=z_{n-1}\tau^2}
w_{n \cdots 1}(z_{1},\cdots,z_{n}) \right) \otimes
w_{\alpha_{n+1} \cdots \alpha_N}(z_{n+1},\cdots,z_N),
\end{array}
\end{equation}
for $\alpha_i =n+1-i~ (1 \leq i \leq n)$.
By combining (\ref{R-goog}) and (\ref{last}) and using (\ref{Chq})
we have the useful expression
\begin{equation}
\begin{array}{rcl}
&&\displaystyle{\rm Res}_{z^{(n-1)}_m =z^{(n)}_m \tau^2 } \cdots
  {\rm Res}_{z^{(1)}_m =z^{(2)}_m \tau^2 }
w_{\alpha}^{\overbrace{n\cdots n}^m
\cdots \hat{i}\cdots \hat{i}
\cdots \overbrace{1\cdots 1}^m \overbrace{i \cdots i}^m}
(z^{(n)}| \cdots | z^{(1)} ) \\
&=& \displaystyle\prod_{k=2}^n \prod_{l=1}^{m-1} \prod_{j=n+1-k}^{n-1}
    b(z_m^{(n)}\tau^{2j}/z_l^{(k)})\prod
    _{\alpha_{mk+l}<\alpha_{jm} 
      \atop 1\leq l<m} b(z_l^{(n-k)}/z_m^{(j)}) \\
& \times & 
\displaystyle (z_m^{(n)})^{n-1} 
(-\tau)^{i-1} 
(\tau^2-1)^{n-1}\tau^{(n-1)^2}[n-1]! \\ 
& \times & 
\displaystyle v^{(n \cdots 2 1)} \otimes 
w_{\alpha'}^{\overbrace{n\cdots n}^{m-1}
\cdots \hat{i}\cdots \hat{i}
\cdots \overbrace{1\cdots 1}^{m-1} 
\overbrace{i \cdots i}^{m-1}}
(z'^{(n)}| \cdots | z'^{(1)} ). 
\end{array}
\end{equation}
for $\alpha'=(\alpha_1, \cdots , \hat{\alpha_{m}}, \cdots ,
\hat{\alpha_{2m}}, \cdots , \alpha_{nm-1} )$ and
$\alpha_{mi}=n+1-i~ (1\leq i \leq n)$.
Here we use the abbreviation 
$z'^{(i)}=(z^{(i)}_1 , \cdots , z^{(i)}_{m-1})$. 

~

\section{Main theorem}

Now we present the main theorem of the present paper.
In what follows we use the abbreviation
$$
\begin{array}{rcl}
z^{(j)}&=&(z_1^{(j)}, \cdots, z_{m-1}^{(j)}, z_m^{(j)})
        =(z'^{(j)}, z_m^{(j)}), \\
z^{(j)}\tau^{\pm 1}&=&
            (z_1^{(j)}\tau^{\pm 1}, \cdots,
             z_{m-1}^{(j)}\tau^{\pm 1}, z_m^{(j)}\tau^{\pm 1})
           =(z'^{(j)}\tau^{\pm 1}, z_m^{(j)}\tau^{\pm 1}), \\
x&=&(x_1 , \cdots , x_M).
\end{array}
$$
The polynomial $\Delta ^{(m)}$ in (\ref{def:F})
is given by
\begin{equation}
\Delta^{(m)}(x| z^{(n)} | \cdots |
z^{(2)} | z^{(1)})
=\det \left( A_{\lambda }^{(m)}(x_{\mu}| z^{(n)} | \cdots |
z^{(2)} | z^{(1)}) \right) _{1\leq \lambda , \mu \leq M}
\label{determ}
\end{equation}
where $M=M_m=(n-1)m-1, N=mn$.
The entries of the $M\times M$ matrix $A^{(m)}$
is defined as follows.
Let us introduce the polynomial
$$
f_{\lambda}^{(N)}(y|z_1, \cdots , z_N)=
\sum_{\kappa =0}^{\lambda -1} (-1)^{\kappa}
((y\tau)^{\lambda -\kappa}-(y\tau^{-1})^{\lambda -\kappa})
\sigma_{\kappa}(z_1, \cdots , z_N),
$$
where $\sigma_{\kappa }(z_1,\cdots,z_n)$ denotes
the $\kappa $-th elementary symmetric polynomials:
$$
\prod_{j=1}^n(t+z_j)=
\sum_{\kappa =0}^n \sigma_{\kappa }(z_1,\cdots,z_n)t^{n-\kappa}.
$$
Note that for $\alpha >0$
\begin{equation}
f_{N+\alpha}^{(N)}(y|z_1,\cdots,z_N)=
y^{\alpha} \left\{ \tau^{\alpha} \prod_{j=1}^{N}(y\tau -z_j )-
\tau^{-\alpha} \prod_{j=1}^{N}(y\tau^{-1} -z_j ) \right\}.
\label{re:f}
\end{equation}
Define the polynomial
\begin{equation}
\begin{array}{cl}
& A_{\lambda }^{(m)}
(x| z^{(n)} | \cdots | z^{(1)} )
= \displaystyle\sum_{k=1}^{n}\prod_{j=1}^{m} (x-z_j^{(k)}\tau^{2k-n-1})
\times \\
& \times
\displaystyle\frac{1}{x}
f_{\lambda }^{((n-1)m)} (x\tau^{n+1-2k}|
z^{(n)}\tau , \cdots , z^{(k+1)}\tau
\stackrel{k}{\hat{~}} z^{(k-1)}\tau^{-1} , \cdots ,z^{(1)}\tau ^{-1} ).
\label{defA}
\end{array}
\end{equation}
This is a homogeneous polynomial of degree $m+\lambda-1$,
symmetric with respect to $z^{(k)}$'s for each $k$, separately.
By the construction (\ref{determ}) and (\ref{defA}), 
$\Delta ^{(m)}$ is a homogeneous polynomial of degree $M_m m+
M_m (M_m +1)/2$ with correct symmetries.
For $n=3$ , it reads as
\[
\begin{array}{cl}
A_{\lambda}^{(m)}(x|z^{(3)}|z^{(2)}|z^{(1)})
&=\displaystyle\prod_{j=1}^m(x-z_j^{(3)}\tau^2)\frac{1}{x}
f_{\lambda}^{(2m)}(x\tau^{-2}|z^{(2)}\tau^{-1},z^{(1)}\tau^{-1})
\\
&+\displaystyle\prod_{j=1}^m(x-z_j^{(2)})\frac{1}{x}
f_{\lambda}^{(2m)}(x|z^{(3)}\tau,z^{(1)}\tau^{-1})
\\
&+\displaystyle\prod_{j=1}^m(x-z_j^{(1)}\tau^{-2})\frac{1}{x}
f_{\lambda}^{(2m)}(x\tau^2|z^{(3)}\tau,z^{(2)}\tau).
\end{array}
\]

The following is the main theorem of this paper.
\begin{thm}~~~~The integral formula
$$
G(z_1,\cdots,z_N)=\sum_\alpha w_\alpha(z_1,\cdots,z_N)
H\bigl(\{z_j\}_{j\in J^{\alpha}_n}\,|\cdots |\,\{z_j\}_{j\in
J^{\alpha}_1}\bigr)
\prod_{\alpha_i < \alpha_j}
{1 \over b(z_i/z_j)}
$$
satisfies (\ref{eqn:R-symm}) and (\ref{eqn:cyc})
with $(\delta_1,\delta_2,\cdots,\delta_n)=
(\tau^{-(n-1)m},\tau^{-(n-1)m-2},
\cdots,\tau^{-(n-1)(m+2)})$, 
where $H$ is defined by 
(\ref{DefH}--\ref{def:F}) and $w_\alpha(z_1,\cdots,z_N)$
is defined by (\ref{eqn:defT}--\ref{eqn:defw}), respectively. 
\label{mainthm}
\end{thm}

~

First of all, we would like to notice the following lemmas:

\begin{lem}~~~~The following holds: 
$$
(S_{M N}\Delta^{(m)})(z^{(n)} | 
\cdots | z^{(2)} | z'^{(1)}, z_0 \tau^{2n} ) =
(S_{M N}\tilde{\Delta}^{(m)})
(x| z^{(n)}| \cdots | z^{(2)} | z'^{(1)}, z_0 ), 
$$
where
$$
\tilde{\Delta}^{(m)}(x| z^{(n)} | 
\cdots | z^{(2)} | z'^{(1)}, z_0 ) =
\det \left( \tilde{A_{\lambda}}^{(m)}
(x_{\mu}| z^{(n)} | \cdots | 
z^{(2)} | z'^{(1)}, z_0 ) \right)
_{1\leq \lambda,\mu \leq n-2},
$$
and
$$
\begin{array}{cl}
&\tilde{A}^{(m)}_{\lambda}
(x|z^{(n)} | \cdots | z^{(2)} | z'^{(1)}, z_0 ) =
\displaystyle\frac{x-z_0 \tau^{n-1}}{x-z_0 \tau^{n+1}}
          A^{(m)}_{\lambda}(x|z^{(n)} | 
\cdots | z^{(2)} | z'^{(1)}, z_0 \tau^{2n}) \\
+ & \displaystyle \frac{z_0 \tau^{n+1} }{x}
\left\{ \tau^{-2} \prod_{j=1}^{N-1}
        \frac{x\tau^{n-1}-z_j  }{z_0 \tau^{2} - z_j }
-\frac{x-z_0 \tau^{n-1}}{x-z_0 \tau^{n+1}}
\prod_{j=1}^{N-1} 
\frac{x\tau ^{-n+1}-z_j }{z_0 \tau ^{2} -z_j }
\right\} \\
\times &  A^{(m)}_{\lambda}(z_0\tau^{n+1}|
z^{(n)} | \cdots | z^{(2)} | z'^{(1)}, z_0 \tau^{2n}). 
\end{array}
$$
\end{lem}

It can be proved in a similar manner as in the case $n=2$,
see \cite{JKMQ}.

~

To prove Theorem \ref{mainthm} 
it is enough to show (\ref{eqn:cyclic}) for
$n$ cases; i.e.,
\begin{equation}
\begin{array}{cl}
& G^{\overbrace{n\cdots n}^m
\cdots \hat{i}\cdots \hat{i}
\cdots \overbrace{1\cdots 1}^m \overbrace{i \cdots i}^m}
( z^{(n)}| \cdots | z'^{(1)}, z_0 \tau^{2n}) \\ 
= & 
\delta_{i}~
G^{i \overbrace{n\cdots n}^m
\cdots \hat{i}\cdots \hat{i}
\cdots \overbrace{1\cdots 1}^m \overbrace{i \cdots i}^{m-1}}
( z_0 | z^{(n)}| \cdots | z'^{(1)}), \label{i=1, n}
\end{array}
\end{equation}
for $i=1, \cdots , n$.

Let $z^{(k)}_j =z_{(n-k)m+j}$ for $n\geq k \geq 2$ and
$z'^{(1)}_j =z_{(n-1)m+j}$.
Let us define $J_i (j_1 , \cdots , j_{i-1})$ recursively as follows:
\begin{equation}
\begin{array}{rcl}
J_1 & = & \{0, 1, \cdots , m \} \ni j_1, \\
J_2 (j_1) & = & \{ j_1 , m+1, \cdots , 2m \} \ni j_2, \\
& \vdots & \\
J_{n-2} (j_1 , \cdots , j_{n-3} )
& = & \{ j_{n-3}, (n-3)m+1, \cdots , (n-2)m \} \ni j_{n-2} \\
J_{n-1} (j_1 , \cdots , j_{n-2} )
& = & \{ j_{n-2}, (n-2)m+1, \cdots , (n-1)m \} \ni j_{n-1} \\
\end{array}
\end{equation}
Set
$$
\begin{array}{cl}
& \varphi ^{(m)} (j_1 , \cdots , j_{n-1}) 
= 
\displaystyle\prod_{k_1 \in J_1 \backslash \{ j_1 \} }
\frac{z_{k_1}-z_0 \tau^2 }{(z_{j_1}-z_{k_1})\tau}
\displaystyle\prod_{k_2 \in J_2 \backslash \{ j_2 \} }
\frac{z_{k_2}-z_{j_1}\tau^2 }{(z_{j_2}-z_{k_2})\tau}
\cdots \\ 
\times & \cdots 
\displaystyle\prod_{k_{n-1} \in J_{n-1} \backslash \{ j_{n-1} \} }
\frac{z_{k_{n-1}}-z_{j_{n-2}} \tau^2 }{(z_{j_{n-1}}-z_{k_{n-1}})\tau}
\displaystyle\prod_{j=1}^{m-1}
\frac{z'^{(1)}_{j}-z_{j_{n-1}} \tau^2 }{(z'^{(1)}_{j}-z_{0})\tau^2 }.
\end{array}
$$
The equation (\ref{i=1, n}) for $i =1$ and $\delta_1 =\tau^{m(1-n)}$
is satisfied if the following holds:
\begin{prop}
\begin{equation}
\begin{array}{cl}
& \displaystyle\sum_{J} \varphi^{(m)}(j_1 , \cdots , j_{n-1})
\Delta ^{(m)} (x| \stackrel{j_1}{\hat{z_{J_1}}}|
\cdots | \stackrel{j_{n-1}}{\hat{z_{J_{n-1}}}} | z_{j_{n-1}}, z'^{(1)}) \\
= & (-\tau)^{m(1-n)}
\displaystyle\prod_{k=2}^n \prod_{j=1}^m
\frac{z^{(k)}_j -z_0 \tau^2 }{z^{(k)}_j -z_0 \tau^{2n-2}}
\tilde{\Delta} (x| z^{(n)}| \cdots | z^{(2)} | z'^{(1)}, z_0 ).
\end{array}
\label{eqn:e=1}
\end{equation}
\label{e=1}
\end{prop}
In section 4 we prove Theorem \ref{mainthm} for $m=1$. 
In section 5 we verify Proposition \ref{e=1} and also show that
(\ref{i=1, n}) for $i\neq 1$ reduces to Proposition \ref{e=1}, 
which imply Theorem \ref{mainthm} for any $m$. 

~

\section{The $m=1$ case}

In this section we prove Theorem \ref{mainthm} for $m=1$.
Notice that for $m=1$
$A_{\lambda }^{(m)} (x| z)$ coincides with
\begin{equation}
A_{\lambda}^{(1)}(x|z_1,\cdots,z_n)
=\sum_{\kappa =0}^{\lambda}
(-1)^{\kappa } x^{\lambda-\kappa }
(\tau^{n(\lambda-\kappa )+\kappa }-\tau^{-n(\lambda-\kappa )-\kappa })
\sigma_{\kappa }(z_1,\cdots,z_n).
\end{equation}
and that it
is linear with respect to $z$'s.
In this case $z_i=z_1^{(n+1-i)}$.
In this section we use the abbreviations
$\Delta^{(1)}=\Delta, \tilde{\Delta}^{(1)}=\tilde{\Delta},
A_{\lambda}^{(1)}=A_{\lambda} $.
Since the polynomial $A_{\lambda}(x|z_1, \cdots, z_n)$
is symmetric with respect to the variable $(z_1, \cdots, z_n)$,
by using (\ref{eqn:Rsym}) we obtain
\begin{equation}
\begin{array}{rcl}
G(z_1,\cdots,z_n)^{n \cdots \hat{i} \cdots 1 i }
&=&(-\tau)^{1-i}H(z_1,\cdots,z_n), \\
G(z_1,\cdots,z_n)^{i n \cdots \hat{i} \cdots 1}
&=&(-\tau)^{i-1}H(z_1,\cdots,z_n).  \label{G-symm}
\end{array}
\end{equation}
Thus
\begin{equation}
G(z_2,\cdots,z_n,z_1\tau^{2n})^{n \cdots 21}=
\delta_1 G(z_1,\cdots,z_n)^{1n\cdots2}, ~~~~
\delta_1 =\tau^{1-n}, \label{i=1}
\end{equation}
implies
\begin{equation}
G(z_2,\cdots,z_n,z_1\tau^{2n})^{n \cdots \hat{i} \cdots 1 i }=
\delta_i G(z_1,\cdots,z_n)^{i n \cdots \hat{i} \cdots 1}, ~~~~
\delta_i =\delta_1 \tau^{2-2i}.
\end{equation}
Consequently, 
to prove Theorem \ref{mainthm} for $m=1$, 
it is enough to show (\ref{i=1}). 

Now we prepare the following lemmas.
\begin{lem}
\begin{equation}
\tilde{\Delta}(x_1, \cdots, x_{n-2}|z_1, \cdots, z_n)|_{z_j=
z_1\tau^{2n-2}}=0,~~~(j=2,3,\cdots,n).
\label{eqn:m=1zero}
\end{equation}
\label{lem:vanish}
\end{lem}
{\sl Proof}~~~~Note that
$$
\begin{array}{cl}
&\displaystyle\sum_{\lambda=0}^{n-3}
z_1^{\lambda}A_{n-2-\lambda}
(x|z_1\tau^{-1},z_1\tau,z_3,\cdots,z_n) \\
=&\displaystyle\frac{1}{x}
\left\{ (x\tau^{n-2}-z_1)\prod_{j=3}^n(x-z_j\tau^{-n+1})
  -(x\tau^{-n+2}-z_1)\prod_{j=3}^n(x-z_j\tau^{n-1})\right\}.
\end{array}
$$
Hence we obtain the linear dependence
$$
\sum_{\lambda=0}^{n-3}
(z_1\tau^{2n-1})^{\lambda}\tilde{A}_{n-2-\lambda}
(x|z_1,z_1\tau^{2n-2},z_3,\cdots,z_n)=0,
$$
which implies (\ref{eqn:m=1zero}).~~~~$\Box$

~

\begin{lem}
\begin{equation}
\begin{array}{cl}
&\displaystyle\prod_{\mu=2}^{n-2}(x_\mu-z_1\tau)
\Delta(z_1\tau^{-1}, x_2, \cdots, x_{n-2}|z_1\tau^{-n}, z_2, \cdots, z_n)
\\
=&\displaystyle\prod_{\mu=2}^{n-2}(x_\mu-z_1\tau^{-1})
\Delta(z_1\tau, x_2, \cdots, x_{n-2}|z_1\tau^n, z_2, \cdots, z_n).
\label{eqn:m=1,2}
\end{array}
\end{equation}
\label{lem:CC'}
\end{lem}

{\sl Proof}~~~~Let us first show that
\begin{equation}
\det(C_{\lambda,\mu})_{1\leq, \mu, \lambda \leq n-2}=
\det(C'_{\lambda,\mu})_{1\leq, \mu, \lambda \leq n-2}, \label{C=C'}
\end{equation}
where
$$
C_{\lambda \mu }=\left\{
\begin{array}{ll}
\displaystyle A_{\lambda }(z_1\tau^{-1}|z_1\tau^{-n}, z_2, \cdots, z_n), &
\mbox{if $\mu =1$,}\\
(x_{\mu}-z_1\tau)A_{\lambda}(x_{\mu}|z_1\tau^{-n}, z_2, \cdots, z_n),
& \mbox{if $\mu \neq 1$, }
\end{array}
\right.
$$
$$
C'_{\lambda \mu }=\left\{
\begin{array}{ll}
\displaystyle A_{\lambda }(z_1\tau|z_1\tau^n, z_2, \cdots, z_n), &
\mbox{if $\mu =1$,}\\
(x_{\mu}-z_1\tau^{-1})A_{\lambda}(x_{\mu}|z_1\tau^n, z_2, \cdots, z_n).
& \mbox{if $\mu \neq 1$. }
\end{array}
\right.
$$
Perform the following elementary transformations
to the matrix $(C_{\lambda \mu})_{1\leq \lambda,\mu\leq n-2}$
$$
\begin{array}{l}
(1) (\mbox{$i$th row})-z_1\tau^{n-1}(\mbox{$(i-1)$st row}), ~~
(i=2, \cdots, n-2),
\\
(2) (\mbox{$i$th column})+z_1\tau(\mbox{$(i-1)$st column}), ~~
(i=2, \cdots, n-2),
\end{array}
$$
and to the matrix $(C'_{\lambda \mu})_{1\leq \lambda, \mu\leq n-2}$
$$
\begin{array}{l}
(1) (\mbox{$i$th row})-z_1\tau^{-n+1}(\mbox{$(i-1)$st row}), ~~
(i=2, \cdots, n-2),
\\
(2) (\mbox{$i$th column})+z_1\tau^{-1}(\mbox{$(i-1)$st column}), ~~
(i=2, \cdots, n-2).
\nonumber
\end{array}
$$
Then we have (\ref{C=C'}) and therefore (\ref{eqn:m=1,2}).
~~~~$\Box$

~

Because the polynomial $A_{\lambda}(z_1,\cdots,z_n)$ is symmetric with
respect to $(z_1,\cdots,z_n)$, 
Proposition \ref{e=1} holds 
if the following Proposition holds.

~

\begin{prop}
$$
\tilde{\Delta}(x_1,\cdots,x_{n-2}|z_1,\cdots,z_n)
=\prod_{j=2}^n\frac{z_j-z_1\tau^{2n-2}}{z_j-z_1\tau^2}
\Delta(x_1, \cdots, x_{n-2}|z_1, \cdots, z_n).
$$
\end{prop}

{\sl Proof}~~~~
Thanks to the linearity of the determinant we have
\begin{equation}
\begin{array}{cl}
&\tilde{\Delta}(x_1, \cdots, x_{n-2}|z_1, \cdots, z_n) \\
=&\displaystyle\prod_{\mu =1}^{n-2}
   \frac{x_{\mu }-z_1 \tau^{n-1} }{x_{\mu }-z_1 \tau ^{n+1} }
   \Delta(x_1, \cdots, x_{n-2}|z_1, \cdots, z_n) \\
+&\displaystyle\sum_{\nu =1 }^{n-2}
\displaystyle\prod_{\mu =1 \atop \mu \neq \nu }^{n-2}
   \frac{x_{\mu}-z_1\tau^{n-1}}{x_{\mu }-z_1 \tau ^{n+1} }
   g(x_{\nu}|z_1|z_2, \cdots, z_n)\\ 
\times & 
\Delta(x_1, \cdots, 
\stackrel{\nu}{\hat{z_1 \tau^{n+1}}} \cdots x_{n-2}|
z_1\tau^{2n}, z_2 , \cdots, z_n),
\end{array}
\label{n-fold} 
\end{equation}
where
$$
g(x|z_1|z_2, \cdots, z_n)=
\displaystyle\frac{z_1 \tau^{n+1} }{x}
\left\{ \tau^{-2} \prod_{j=2}^n
        \frac{x\tau^{n-1}-z_j  }{z_1 \tau^{2} - z_j }
-\frac{x-z_1 \tau^{n-1}}{x-z_1 \tau^{n+1}}
\prod_{j=2}^n \frac{x\tau ^{-n+1}-z_j }{z_1 \tau ^{2} -z_j }\right\}.
$$
Hence we get
\begin{equation}
\begin{array}{cl}
&\tilde{\Delta}(z_1\tau^{n-1}, x_2, \cdots, x_{n-2}|z_1, \cdots, z_n) \\
=& \displaystyle\prod_{j=2}^{n}\frac{z_j-z_1\tau^{2n-2}}{z_j-z_1\tau^2}
\prod_{\mu=2}^{n-2}\frac{x_\mu-z_1\tau^{n-1}}{x_\mu-z_1\tau^{n+1}}
\Delta(z_1\tau^{n+1}, x_2, \cdots, x_{n-2}|z_1\tau^{2n}, \cdots, z_n).
\label{m=3,4}
\end{array}
\end{equation}
{}From Lemma \ref{lem:CC'} and (\ref{m=3,4}), we obtain
\begin{equation}
\tilde{\Delta}(z_1\tau^{n-1}, x_2, \cdots, x_{n-2}|z_1, \cdots, z_n)
=\prod_{j=2}^{n}\frac{z_j-z_1\tau^{2n-2}}{z_j-z_1\tau^2}
\Delta(z_1\tau^{n-1}, x_2, \cdots, x_{n-2}|z_1, \cdots, z_n).
\end{equation}

The polynomials $\tilde{\Delta}$ and $\Delta$ have the same factor
$\prod_{1 \leq \mu \leq \nu \leq n-2} 
(x_{\mu}-x_{\nu})$.
Furthermore, the degrees of $\tilde{\Delta}$ and $\Delta$ with
respect to $x_{\mu}$ are at most $(n-2)$.
Thus we get
\begin{equation}
\begin{array}{cl}
&\displaystyle\tilde{\Delta}(x_1,x_2,\cdots,x_{n-2}|z_1,\cdots,z_n)
-\prod_{j=2}^{n}\frac{z_j-z_1\tau^{2n-2}}{z_j-z_1\tau^{2}}
\Delta(z_1\tau^{n-1},x_2,\cdots,x_{n-2}|z_1,\cdots,z_n)
\\
=& \displaystyle c(z_1,\cdots,z_n)
\prod_{1\leq \mu<\nu \leq n-2}(x_\mu-x_\nu)
\prod_{\mu=1}^{n-2}(x_\mu-z_1\tau^{n-1}),
\label{eqn:up}
\end{array}
\end{equation}
where $c(z_1,\cdots,z_n)$ is homogeneous rational function of
$(z_1,\cdots,z_n)$ whose total degree is $0$.
{}From Lemma \ref{lem:vanish} 
$c$ has zeros at $z_j =z_1 \tau^{2n-2}$,
and may have poles at only $z_j = z_1 \tau^2 $.
Thus $c$ must have the form
$$
c(z_1, \cdots, z_n)=s\prod_{j=2}^n
\frac{z_j-z_1\tau^{2n-2}}{z_j-z_1\tau^2},~~~~ s\in \bc.
$$
By comparing both sides of (\ref{eqn:up}) at $z_1=0$, we obtain $s=0$.
~~~~$\Box$

Therefore, by taking into account 
the symmetry (\ref{G-symm}), 
Theorem \ref{mainthm} for $m=1$ 
was proved. 

~

\section{General case}
In this section we prove Theorem \ref{mainthm} for general case.
Let $I_l =\{
(\gamma_1 , \cdots , \gamma_l ) | n \geq \gamma_1 > \cdots \gamma_l
\geq 1 \}$
for $1\leq l \leq n-1$.
Fix nonnegative integers $m(\gamma_1 , \cdots , \gamma_l )$
for $1\leq l \leq n-1$ and set
$$
N_k = \sum_{l=1}^{n-1} \sum_{(\gamma_1 , \cdots , \gamma_l ) \in I_l
\atop \gamma_j \neq k} m(\gamma_1 , \cdots , \gamma_l ).
$$
In what follows
we often use the abbreviation $\gamma=(\gamma_1,\cdots,\gamma_l)$.
Define the polynomial with respect to variables
$x$ and $\zeta^{(\gamma)}=(\zeta_1^{(\gamma)}, \cdots ,
\zeta_{m(\gamma)}^{(\gamma)})$
by
\begin{equation}
\begin{array}{cl}
&A_{\lambda}
(x|\zeta^{(n)}|\cdots |\zeta^{(1)}|
   \zeta^{(nn-1)}|\cdots |\zeta^{(21)}|\cdots
  |\zeta^{(\gamma_1  \cdots  \gamma_l )}|\cdots
  |\zeta^{(n-1 \cdots 2 1)}) \\
=&\displaystyle\sum_{k=1}^n \prod_{l=1}^{n-1}
                \prod_{(\gamma)\in I_l,
                        \atop \gamma_i=k}
                \prod_{j=1}^{m(\gamma)}
                (x-\zeta_j^{(\gamma)}\tau^{2k-n-3+2i}) \\
\times & \displaystyle\frac{1}{x}
f_{\lambda }^{(N_k)}(x\tau^{n+1-2k}|
   \bigcup_{l=1}^{n-1}
   \bigcup _{(\gamma)\in I_l,
        \atop \gamma_i>k>\gamma_{i+1}}
   \bigcup_{j=1}^{m(\gamma)}
   \zeta_j^{(\gamma)}\tau^{2i-1}~). 
\end{array}
\label{eqn:zeta}
\end{equation}
For $n=3$ it reads as
$$
\begin{array}{cl}
& x A_{\lambda} (x| \zeta^{(3)}| \zeta^{(2)}| \zeta^{(1)} |
\zeta^{(32)}| \zeta^{(31)}| \zeta^{(21)} ) \\
= & \displaystyle \prod_{j=1}^{m(3)} (x-\zeta ^{(3)}_j \tau^2 )
                  \prod_{j=1}^{m(3, 2)} (x-\zeta ^{(32)}_j \tau^2 )
                  \prod_{j=1}^{m(3, 1)} (x-\zeta ^{(31)}_j \tau^2 )
\displaystyle f_{\lambda }^{(N_3)}
(x\tau^{-2} | \zeta^{(2)}\tau^{-1}, 
\zeta^{(1)}\tau^{-1}, \zeta^{(21)}\tau^{-1}) \\
+ & \displaystyle \prod_{j=1}^{m(2)} (x-\zeta ^{(2)}_j )
                  \prod_{j=1}^{m(3, 2)} (x-\zeta ^{(32)}_j \tau^2 )
                  \prod_{j=1}^{m(2, 1)} (x-\zeta ^{(21)}_j )
\displaystyle f_{\lambda }^{(N_2)}
(x | \zeta^{(3)}\tau, \zeta^{(1)}\tau^{-1}, \zeta^{(31)}\tau ) \\
+ & \displaystyle 
\prod_{j=1}^{m(1)} (x-\zeta ^{(1)}_j \tau^{-2} )
\prod_{j=1}^{m(3, 1)} (x-\zeta ^{(31)}_j )
\prod_{j=1}^{m(2, 1)} (x-\zeta ^{(21)}_j )
\displaystyle f_{\lambda }^{(N_1)}
(x\tau^{2} | \zeta^{(3)}\tau, \zeta^{(2)}\tau, 
 \zeta^{(32)}\tau^{3}).
\end{array}
$$

It follows from the recursion relation
$$
f_{\lambda}^{(N+1)}
    (y|z_1,\cdots,z_N, a)
=f_{\lambda}^{(N)}(y|z_1,\cdots,z_N)-
af_{\lambda-1}^{(N)}(y|z_1,\cdots,z_N),
$$
that
the polynomial $A_{\lambda}^{(m)}$ satisfies
\begin{equation}
\begin{array}{cl}
&\displaystyle A_{\lambda }^{(m)}(x|z^{(n)}|\cdots|z^{(1)})|
  _{\bigcup _{l=1}^{n-1} \bigcup _{j=1}^l z^{(\gamma_j)}
     =\zeta ^{(\gamma_1 \cdots \gamma_l)}\tau^{2j-2}}
\\
=& \displaystyle\sum_{\rho =0}^L(-1)^{\rho}
    \sigma _{\rho}(\bigcup _{l=2}^{n-1}
    \bigcup_{\gamma \in I_l} \bigcup _{j=1}^{l-1}
    \zeta ^{(\gamma)}_j \tau^{2j-1})
    A_{\lambda-\rho}(x|\zeta^{(n)}|\cdots|\zeta^{(1)}|\cdots|
        \zeta^{(n-1 \cdots 2 1)}), 
\end{array}
\label{restofA}
\end{equation}
where
$$
\sum_{l=1}^{n-1}\sum_{(\gamma_1 ,\cdots , \gamma_l)\in I_l 
                       \atop \gamma_j=k}
m(\gamma_1,\cdots,\gamma_l)=m,~~~~(k=1,2,\cdots,n),
$$
and
$$
L=\sum_{l=2}^{n-1}(l-1)\sum_{\gamma \in I_l} m(\gamma).
$$
For $n=3$ it reads as
$$
\begin{array}{cl}
&\displaystyle A_{\lambda }^{(m)}(x|z^{(3)}|z^{(2)}|z^{(1)})|
  _{\bigcup _{l=1}^{2} \bigcup _{j=1}^l z^{(\gamma_j)}
     =\zeta ^{(\gamma_1 \cdots \gamma_l)}\tau^{2j-2}}
\\
= & \displaystyle\sum_{\rho =0}^{L} 
(-1)^{\rho} \sigma_{\rho } 
(\zeta^{(32)}\tau , \zeta^{(31)}\tau , 
\zeta^{(21)}\tau )
A_{\lambda -\rho }(x| 
\zeta^{(3)}\tau , \zeta^{(2)}\tau , \zeta^{(1)}\tau , 
\zeta^{(32)}\tau , \zeta^{(31)}\tau , \zeta^{(21)}\tau ). 
\end{array}
$$

Define the polynomial for positive integer $\alpha$
$$
\begin{array}{cl}
&\displaystyle h_{\alpha}^{(m|L)}
    (x|\zeta^{(n)}|\cdots |\zeta^{(1)}|\zeta^{(n n-1)}|
    \cdots |\zeta^{(\delta_1 \cdots \delta_l)}|\cdots
    |\zeta^{(n-1 \cdots 2 1)}) \\
=& \displaystyle x^{\alpha -2} \left\{
\tau^{n(\alpha-1)} \prod_{l=1}^{n-1}
    \prod_{\gamma \in I_l} \prod_{j=1}^{m(\gamma)}
    (x\tau^{n-l(\gamma)}-\zeta_j^{
     (\gamma)}\tau^{l(\gamma)-1}) \right. \\
- & \displaystyle 
\left. 
\tau^{-n(\alpha-1)} \prod_{l=1}^{n-1}
    \prod_{\gamma \in I_l} \prod_{j=1}^{m(\gamma)}
    (x\tau^{-n+l}-\zeta_j^{(\gamma)}\tau^{l(\gamma)-1}) \right\}.
\end{array}
$$
One of important observations is as follows:

~

\begin{prop}~~~~
For $~m(\gamma_1,\gamma_2,\cdots,\gamma_l)$ such that
$\sum_{l=1}^{n-1}\sum_{(\gamma_1,\cdots,\gamma_l)\in I_l,\gamma_j=k}
m(\gamma_1,\cdots,\gamma_l)=m,(k=1,2,\cdots,n)$,
the polynomial $\Delta^{(m)}$ satisfies the following relation.
\begin{equation}
\begin{array}{cl}
&\displaystyle\Delta^{(m)}(x_1,\cdots,x_M|z^{(n)}|z^{(n-1)}|
   \cdots|z^{(1)})
   |_{\bigcup _{l=1}^{n-1} \bigcup _{j=1}^l z^{(\gamma_j)}
     =\zeta ^{(\gamma_1 \cdots \gamma_l)}\tau^{2j-2}} \\
=& \displaystyle\sum_{1\leq \mu_1 <\cdots <\mu_L\leq M}
(-1)^{\sum_{i=1}^{L} M-L +i + \mu_i }
\det \left( h_{\alpha-1}^{(m)}
(x_{\mu_i}|\zeta^{(n)}|\cdots|\zeta^{(1)}|\cdots
|\zeta^{(n-1 \cdots 2 1)} \right)_{1\leq i, \alpha \leq L} \\
\times & \displaystyle\Delta(\stackrel{x_{\mu}}{\hat{x}} |
   \zeta^{(n)}|\cdots|\zeta^{(1)}|\cdots|\zeta^{(n-1\cdots 2 1)}).
\end{array}
\label{parec:A}
\end{equation}
Here
\begin{equation}
\begin{array}{cl}
&\Delta(y_1, \cdots, y_{M-L}|
 \zeta^{(n)}|\cdots|\zeta^{(1)}|
\cdots|\zeta^{(n-1 \cdots 2 1)}) \\
= & 
\det \left( A_{\lambda}(y_{\mu}|\zeta^{(n)}|\cdots |\zeta^{(1)}|
                \zeta^{(nn-1)}|\cdots |\zeta^{(21)}|\cdots
                |\zeta^{(n-1 \cdots 1)} \right)_{ 
1\leq \lambda ,\mu \leq M-L},
\end{array}
\label{parec:D} 
\end{equation}
and
$$
L=\sum_{l=2}^{n-1}(l-1)\sum_{\gamma \in I_l}m(\gamma),~~M=(n-1)m-1.
$$
\label{parec}
\end{prop}

{\sl Proof}~~~~
This follows from the induction with respect to $L$.
When $L=0$ (\ref{parec:A}) is obvious.
Assume (\ref{parec:A}) for $L$.
After another restriction, say
$\zeta^{(1)}_{m(1)}=\zeta^{(2)}_{m(2)}\tau^2$, 
using (\ref{re:f}) we have
$$
\begin{array}{cl}
& A_{M-L}(x|\zeta^{(n)}|\cdots |\zeta^{(1)}|
                \zeta^{(n n-1)}|\cdots |\zeta^{(2 1)}|\cdots
          |\zeta^{(n-1 \cdots 1)})|_{ 
           \zeta^{(1)}_{m(1)}=\zeta^{(2)}_{m(2)}\tau^2} \\
= &
h^{(m|L+1)}_{1}
    (x|\zeta^{(n)}|\cdots |\zeta^{(1)}|\zeta^{(n n-1)}|
    \cdots |\zeta^{(\delta_1 \cdots \delta_l)}|\cdots
    |\zeta^{(n-1 \cdots 2 1)}).
\end{array}
$$
Furthermore, we have
$$
\begin{array}{cl}
& h^{(m|L)}_{\alpha}(x|\zeta^{(n)}|\cdots |\zeta^{(1)}|
 \cdots |\zeta^{(n-1 \cdots 1)})|_{
 \zeta^{(1)}_{m(1)}=\zeta^{(2)}_{m(2)}\tau^2} \\
= &
h^{(m|L+1)}_{\alpha+1}(x|\zeta^{(n)}|\cdots |\zeta^{(1)}|
                \cdots |\zeta^{(n-1 \cdots 1)}) -
\zeta^{(2)}_{m(2)}\tau
h^{(m|L+1)}_{\alpha}(x|\zeta^{(n)}|\cdots |\zeta^{(1)}|
                \cdots | \zeta^{(n-1 \cdots 1)}).
\end{array}
$$
For other types of restrictions we get similar results.
Thus using (\ref{restofA}) and
performing elementary transformation we have (\ref{parec:A})
for $L+1$. ~~~~$\Box$

~

Thanks to the recursion relation for
$f_{\lambda}^{(N)}$
$$
f_{\lambda}^{(N+n-1)}
    (y|z_1,\cdots,z_N, a\tau, a\tau^{3},\cdots,a\tau^{2n-3})
=\sum_{\rho=0}^{n-1}
    (-1)^{\rho} \sigma_{\rho}(a\tau, a\tau^3 , \cdots , a\tau^{2n-3})
f_{\lambda-\rho }^{(N)}(y|z_1,\cdots,z_N), 
$$
the polynomial $A_{\lambda}^{(m)}$ satisfies
the following recursion relation:
\begin{equation}
\begin{array}{cl}
&A_{\lambda}^{(m)}
    (x|z^{(n)}|\cdots|z^{(1)})
    |_{z_m^{(k)}=a\tau^{2n-2k}~~(k=1, \cdots, n)}
=(x-a\tau^{n-1}) \\
\times & \displaystyle\sum_{\rho =0}^{n-1}
(-1)^{\rho} \sigma_{\rho}(a\tau, a\tau^3 , \cdots , a\tau^{2n-3})
 A_{\lambda -\rho}^{(m-1)}(x|z'^{(n)}|\cdots|z'^{(1)}).
\end{array}
\label{m,m-1}
\end{equation}
Note that for $1\leq \alpha \leq n-1$
\begin{equation}
A^{(m-1)}_{M_{m-1}+\alpha }(z'^{(n)}| \cdots | z'^{(1)})
=
h_{\alpha }^{(m-1|0)}(x|z').  \label{def:h}
\end{equation}
By combining (\ref{m,m-1}) and (\ref{def:h})
and using the same argument in the proof for Proposition \ref{parec} ,
one can show

~
\begin{prop}~~~~
The determinant $\Delta^{(m)}$ obeys the following recursion relation.
$$
\begin{array}{cl}
& \displaystyle\Delta^{(m)}(x_1, \cdots, x_M| z^{(n)}|\cdots|
z^{(1)})|_{z_m^{(k)}=a\tau^{2n-2k},(k=1,2,\cdots,n)}
\\
=& \displaystyle\prod_{\mu=1}^M (x_{\mu}-a\tau^{n-1})
        \sum_{1\leq \mu_1<\cdots <\mu_{n-1}\leq M}
        (-1)^{\sum_{i=1}^{n} mn-m-n+i +\mu_i } \label{Aa} \\
\times &
\displaystyle\det\left( h_{\alpha}^{(m-1|0)}(x_{\mu_i}|z')\right) _{
1\leq i, \alpha \leq n-1}
        \Delta^{(m-1)}
        (x\setminus \{x_{\mu}\}|z'),  
\end{array}
$$
where we use the abbreviation
$x_{\mu}=(x_{\mu_1},\cdots,x_{\mu_{n-1}})$. 
\label{Deltarec}
\end{prop}

~

The following are two key theorems:

~

\begin{thm}~~~~Let $P(m)$ be Proposition \ref{e=1} for $m\geq 1$.
Then for $m>1$, $P(m)$ under the restriction
$z^{(n)}_m=a, \cdots , z^{(2)}_m =a\tau^{2n-4},
z^{(1)}_{m-1}=a\tau^{2n-2}$ holds if $P(m-1)$ holds.
\label{thm1}
\end{thm}

{\sl Proof}~~~~Using the relation
$$
\varphi^{(m)}(j_1 , \cdots , j_{n-1}) |_{
z_m^{(k)}=a\tau^{2n-2k},(k=2\cdots n),z_{m-1}^{(1)}=a\tau^{2n-2}}
= (-\tau)^{n-1} \frac{a-z_0 \tau^2 }{a\tau^{2n-2}-z_0 \tau^2}
\varphi^{(m-1)}(j_1 , \cdots , j_{n-1}), 
$$
the LHS of (\ref{eqn:e=1}) 
can be expressed in a recursive way.
Let us turn to the RHS. Note that
\begin{equation}
\begin{array}{cl}
& \tilde{h}^{(m|l)}_{1}(x|z'^{(n)}|\cdots |z'^{(n-l+1)}|
z^{(n-l)}| \cdots | z^{(1)}|\cdots | \zeta^{(n \cdots n-l+1)}=a) \\
= & 
\displaystyle\frac{z_0 
\tau^{2n-2}-a\tau^{2l-4}}{z_0 -a\tau^{2n-2}} 
\displaystyle\frac{1}{x} \left\{
(x\tau^{n-l}-a\tau^{l-1})\prod_{j=0 \atop j\neq km (k=1, \cdots , l)}
(x\tau^{n-1}-z_j) \right. 
\\ 
- & \displaystyle\left. 
(x\tau^{-n+l}-a\tau^{l-1})
\prod_{j=0 \atop j\neq km (k=1, \cdots , l)}
(x\tau^{-n+1}-z_j) \right\}.
\end{array}
\label{itar}
\end{equation}
{}From Proposition \ref{Deltarec} and (\ref{itar})
we obtain
$$
\begin{array}{cl}
&\displaystyle\tilde{\Delta}^{(m)}(x|z^{(n)}|\cdots|z'^{(1)},z_0)|
       _{z_m^{(k)}=a\tau^{2n-2k},(k=2\cdots n),z_{m-1}^{(1)}=a\tau^{2n-2}}
\\
=&\displaystyle\prod_{l=2}^{n}\frac{z_0 \tau^{2n-2}-a\tau^{2l-4}}
                         {z_0 -a \tau^{2l-4}}
   \prod_{\mu=1}^M (x_{\mu}-a\tau^{n-1})
\sum_{1\leq \mu_1 <\cdots <\mu_{n-1}\leq M}
(-1)^{\sum_{i=1}^n mn-m-n+i+\mu_i} \\
\times & \displaystyle \det \left( h_{\alpha}^{(m-1)}(x_{\mu_i}|z'',z_0)
\right)_{1\leq i,\alpha \leq n-1}
\tilde{\Delta}^{(m-1)}(x\setminus \{x_{\mu}\}|z'',z_0),
\end{array}
$$
where we use the abbreviation
$z''=(z'^{(n)}|\cdots|z'^{(2)}|z^{(1)}_1, \cdots, z^{(1)}_{m-2})$.
Thus the claim is verified. ~~~~$\Box$

~
\begin{thm}~~~~Let $P(m)$ be Proposition \ref{e=1} for $m\geq 1$
Then $P(m)$ under the restriction
$z^{(n)}_m=z_0 \tau^{2}, \cdots , z^{(2)}_m =z_0 \tau^{2n-2}$
holds.
\label{thm2}
\end{thm}

{\sl Proof}~~~~Under the restriction in consideration
there is only one nonzero term in the LHS of (\ref{eqn:e=1}), 
and the only nonzero 
coefficient $\varphi^{(m)}$ is
\begin{equation}
\varphi^{(m)}(m, 2m, \cdots , (n-1)m)=
(-\tau)^{m(1-n)} \prod_{j=1}^{m-1}
\frac{z'^{(1)}_j -z_0 \tau^{2n}}{z'^{(1)}_j -z_0 \tau^{2}}.
\label{Lz0}
\end{equation}
Since in the RHS there exists a zero and a pole 
under the restriction, 
we have to set $z^{(n)}_m=a, \cdots , z^{(2)}_m =a\tau^{2n-4}$
and reduce, then we have to set $a=z_0 \tau^2$ 
to evaluate the RHS: 
\begin{equation}
\begin{array}{cl}
&\tilde{\Delta}^{(m)}(x|z^{(n)}|\cdots|z^{(2)}|z'^{(1)},z_0)
   |_{z_m^{(k)}=z_0\tau^{2n+2-2k},(k=2,\cdots,n)}
\\
=&(-\tau^{(n-1)})^{(n-2)}
\displaystyle\prod_{k=2}^n \prod_{j=1}^{m-1}
\frac{z_j^{(k)}-z_0\tau^{2n-2}}{z_j^{(k)}-z_0\tau^2}
\prod_{k=2}^n
\frac{z_j^{(1)}-z_0\tau^{2n}}{z_j^{(1)}-z_0\tau^2}
\\
\times& \displaystyle\prod_{\mu=1}^M(x_{\mu}-z_0\tau^{n-1})
\sum_{1\leq \mu_1 <\cdots <\mu_{n-1}\leq M}
(-1)^{\sum_{i=1}^n mn-m-n+i+\mu_i}
\\
\times &\displaystyle\det 
(h_{\alpha}^{(m-1)}(x_{\mu_i}|z'))_{1\leq i,\alpha \leq n-1}
\Delta^{(m-1)}(x\setminus\{x_{\mu}\}|z').
\end{array}
\label{Rz0}
\end{equation}
Therefore this proposition follows from (\ref{Lz0}) and (\ref{Rz0}).
~~~~$\Box$

~

We now wish to show Proposition \ref{e=1}.
First of all, note that LHS of Proposition \ref{e=1} has no singularity at
$z'^{(1)}=z_0 \tau^{2n-2}$. Thus let us show the following proposition.

~

\begin{prop}
$$
\tilde{\Delta}^{(m)}(x_1,\cdots,x_M|z^{(n)}|
        z^{(n-1)}|\cdots|z'^{(1)},z_0)
   |_{z_j^{(k)}=z_0\tau^{2n-2}}=0,~~(k=2,\cdots,n)
$$
\label{poleless}
\end{prop}

{\sl Proof.}~~~~By the same argument as in 
Proposition \ref{Deltarec},
we have an equation which can be obtained from (\ref{parec:D})
by replacing $A\rightarrow \tilde{A}$ and $\Delta
\rightarrow \tilde{\Delta}$.
Since the last row obtained in this way vanishes,
the claim of this Proposition is verified. ~~~~$\Box$

~

Using a general $m$ analogue of (\ref{n-fold}), we can show

~

\begin{prop}~~~~
$$
\begin{array}{cl}
& \displaystyle\lim_{z_m^{(n)}\to z_0 \tau^2 }
\prod_{k=2}^n \prod_{j=1}^m
\frac{z^{(k)}_j -z_0 \tau^2 }{z^{(k)}_j -z_0 \tau^{2n-2}}
\tilde{\Delta} (x| z^{(n)}| \cdots | z^{(2)} | z'^{(1)}, z_0 ) \\
=&
\displaystyle\sum_{\nu =1 }^{M} \prod_{\mu =1 \atop \mu \neq \nu }^{M}
   \frac{x_{\mu}-z_0\tau^{n-1}}{x_{\mu }-z_0 \tau ^{n+1} }
   g_1 (x_{\nu}| z'^{(n)}| z^{(n-1)}| \cdots | z'^{(1)} | z_0 ) 
\\
\times & 
\Delta(x_1, \cdots, \stackrel{\nu}{\hat{z_0 \tau^{n+1}}} \cdots x_{M}|
z'^{(n)}| z^{(n-1)}| \cdots | z'^{(1)} | z_0 ),
\end{array}
$$
where
$$
g_1 (x| z'^{(n)}| z^{(n-1)}| \cdots | z'^{(1)} | z_0 )
=
\displaystyle\frac{z_0 \tau^{n+1} }{x}
\left\{ \tau^{-2} \prod_{j=1 \atop j\neq m}^N
        \frac{x\tau^{n-1}-z_j  }{z_0 \tau^{2} - z_j }
-\frac{x-z_0 \tau^{n-1}}{x-z_0 \tau^{n+1}}
\prod_{j=1 \atop j\neq m}^N 
\frac{x\tau ^{-n+1}-z_j }{z_0 \tau ^{2} -z_j }\right\}.
$$
\label{rez0}
\end{prop}

~
Now we are in a position to prove Proposition \ref{e=1}.

{\sl Proof of Proposition \ref{e=1}}~~~~
{}From Proposition 5.5 both sides of (\ref{eqn:e=1}) are
homogeneous rational functions of degree
$Mm+M(M-1)/2$ with simple poles located at $z'^{(1)}=z_0\tau^2$.
Taking into account that the
antisymmetry of $x_{\mu}$'s the equation (\ref{eqn:e=1}) 
holds if
both sides of (\ref{eqn:e=1}) coincides at $(n-1)m^2$ points
$z^{(n)}=z_0\tau^2$ and $z^{(a)}=z^{(b)}\tau^2 (a<b)$.
{}From Proposition 5.1 both sides of (\ref{eqn:e=1}) 
after $L$ times restriction
can be expressed as follows:
$$
\begin{array}{rcl}
{\rm LHS|_{restrictions}} & = & 
\sum_{1\leq \mu_1 < \cdots < \mu_L \leq M}
\det(h_{\alpha-1}^{(m)}(x_{\mu_i}|\zeta ))_{1\leq \alpha,i\leq L}
S_L(x \setminus \{x_{\mu}\}), 
\\
{\rm RHS|_{restrictions}} & = & 
\sum_{1\leq \mu_1 < \cdots < \mu_L \leq M}
\det(h_{\alpha-1}^{(m)}(x_{\mu_i}|\zeta ))_{1\leq \alpha,i\leq L}
S_R(x \setminus \{x_{\mu}\}).
\end{array}
$$
Here $S_L(y_1,\cdots,y_{M-L})$
and $S_R(y_1,\cdots,y_{M-L})$ are skew symmetric with respect to
$(y_1,\cdots,y_{M-L})$.
Note that the degrees of $S_L (y_1 , \cdots )$ and
$S_R(y_1 , \cdots )$
with respect to
$y_1 $ are $M-L+m-1$.
Furthermore, $\det(h_{\alpha-1}^{(m)}
(y_1 , y_2, \cdots , y_L |\zeta ))_{1\leq \alpha,j\leq L}
=cy_1^{M+m-1}y_2^{M+m-2}\cdots y_L^{M+m-L}+
\cdots .$
By comparing the coefficients 
of ${\rm LHS|_{restrictions}}$ and ${\rm RHS|_{restrictions}}$
with respect to
$x_1^{M+m-1}x_2^{M+m-2}$
$\cdots x_L^{M+m-L}$,
we conclude that
${\rm LHS|_{restriction}}={\rm RHS|_{restriction}}$
if and only if
\begin{equation}
S_L(x_1,\cdots,x_{M-L})=S_R(x_1,\cdots,x_{M-L}).
\label{eqn:afterL}
\end{equation}
Consequently, taking into account that
the antisymmetry of $x_{\mu}$'s, we have to only
examine ${\rm LHS|_{restriction}}
={\rm RHS|_{restriction}}$ at
$(n-1)m^2-Lm$ planes.
After repeating this procedure
the equation (\ref{eqn:e=1}) 
reduces to Theorem \ref{thm2} and \ref{thm2}.
For simplicity and transparency, we first demonstrate
how this induction scheme does work when $n=3$.

For $3\geq a>b \geq1$, let $m(a,b)$ stand for the number of restrictions
$z^{(b)}=z^{(a)}\tau^2$ and let $0 \leq l \leq 1$
be the number of restriction
$z^{(3)}=z_0\tau^2$.
We denote such restriction by $(l,m(2,1)+m(3,2))$
and introduce a lexicographical order by
$(1,m)>(1,m-1)> \cdots > (1,0) >(0,m)> \cdots > (0,0)$.
The restriction $(l,k)$ is larger than $(l',k')$ if
$(l,k)>(l',k')$.
We wish to show that
$(l, m(2,1)+m(3,2))$ reduces to larger restrictions
and $(1,m)$ reduces to Theorem \ref{thm1} and \ref{thm2}.

{\it 1st step.} Set $(l,m(2,1)+m(3,2))=(1,m)$.
In this case we need $m^2-m(1+m(3,1))$ planes on which
(\ref{eqn:afterL}) holds for $L=m+1+m(3,1)$.
We have $m$ restrictions
which reduce to Theorem \ref{thm2},
and $m(m-1)-m(2,1)m(3,2)$ restrictions which reduce 
to Theorem \ref{thm1}, and 
$(m-1-m(3,1)-m(3,2))(m-1-m(3,1)-m(2,1))$ 
restrictions which reduce 
to previous two restrictions.
Hence we actually have $m^2-m(1+m(3,1))+(m(3,1)+1)^2 $ planes.

{\it 2nd step.}
Set $(l,m(2,1)+m(3,2))=(1,k)$, where $k<m$. In this case we need
$2m^2-m(1+k+m(3,1))$ planes on which (\ref{eqn:afterL}) holds.
We have $m$ restrictions which reduce 
to Theorem \ref{thm2},
$k(m-1)-m(3,2)m(2,1)$
restrictions which reduce to Theorem \ref{thm1},
and $(m-k)(2m-k-2)$ restrictions which reduce to
larger restrictions.
We also have $(m-1-m(3,1)-m(3,2))(m-1-m(3,1)-m(2,1))$
additional restrictions at $z^{(1)}=z^{(3)}\tau^2 $
which reduce to previous three restrictions.
Hence we have $2m^2 -m(1+k+m(3,1))+ (k-m+1+m(3,1)/2)^2 +
m(3,1)(3m(3,1)+4)/4$ planes.

{\it 3rd step.} We can show
in a similar way as in the 2nd step that
this induction scheme does work for $l=0$.

Therefore the equation (\ref{eqn:e=1}) are proved for $n=3$.

Next we show (\ref{eqn:e=1}) for general $n$.

Let $l$ denote the maximal length of restrictions
such that
$z^{(n+1-l)}=z^{(n+2-l)}\tau^{2}=
\cdots =z^{(n)}\tau^{2l-2}=z_0 \tau^{2l} $.
Let $r_{j} (1 \leq j \leq n-2)$ denote the number of
restrictions of type
$z^{(k)}=z^{(k+1)}\tau^2=
\cdots =z^{(k+j)}\tau^{2j}$, where $1 \leq k \leq n-j$.
Set $L=\sum_{j=1}^{n-2} jr_j$.
We introduce the lexicographical order in
$\{ (l,L,r_{n-2},\cdots ,r_1)~|~0\leq L+l\leq (n-2)m+1 \}$
by
$$
\begin{array}{l}
(n-2, (n-2)(m-1)+1 , m-1, 0, \cdots , 0, 1)
> (n-2, (n-2)(m-1)+1 , m-2, 1, 0, \cdots , 0, 2) \\
> \cdots > (0, 1, 0, \cdots , 0, 1) > (0, 0, 0, 0, \cdots , 0).
\end{array}
$$
At the stage of induction of type
$(l,L,r_{n-2},\cdots ,r_1)$, we need more than or equal to
$(n-1)m^2 -m(l+L)$ restriction planes.

{\it Remark 1. } In the expression of (\ref{eqn:zeta})
let us call $z$ which belongs to one of $\zeta^{(i)}$
{\it free $z$}. Then we have at least $m-2$ free $z$'s.

{\it Remark 2. } There are restrictions which are not counted
by $r_j (j=1, \cdots , n-2)$; e.g., restrictions
counted by $m(a,b)$, where $a-b>1$. Let us call such a restriction
a {\it bad restriction}.
Even if we need less than or equal to
$m$ bad restrictions to obtain $(n-1)m^2 -m(l+L)$ restriction planes
for $(l,L,r_{n-2}, \cdots , r_1)$,
our induction scheme does work. Because after a bad restriction,
we have still as the same number of not bad restrictions as
that for $(l,L,r_{n-2}, \cdots , r_1)$, while we need
only $(n-1)m^2 -m(l+L)-m$ restriction planes.

{\it 1st step. } Set
$(l,L,r_{n-2},\cdots ,r_1)=
(n-2, (n-2)(m-1)+1 , m-1, 0, \cdots , 0, 1)$. Then we need
$m^2 -m$ planes on which (\ref{eqn:afterL}) holds.
Let us first consider $m(n, \cdots ,2)=m-1,
m(2,1)=1, m(1)=m-2$, and the other $m(\gamma)=0$.
We have $m$ planes which reduce to
Theorem \ref{thm2}, and $(m-1)^2 $ planes which reduce to
Theorem \ref{thm1}. Thus we have actually $m^2 -m +1$ planes.
Next consider $m(n, \cdots ,2)=m-2$ and $m(n-1, \cdots ,1)=1$.
In this case we have $m$ planes which reduce to
Theorem \ref{thm2}, and $(m-1)^2 -(m-2)$ planes which reduce to
Theorem \ref{thm1}. We also have $m-2$ bad restrictions.
Hence from Remark 2 this case reduces to Theorems \ref{thm1} and \ref{thm2}.
In general when we replace
$(m(n, \cdots ,2), m(n-1, \cdots ,1))=(m-p,p-1)$ by
$(m(n, \cdots ,2), m(n-1, \cdots ,1))=(m-1-p,p)$,
we need $m-2p$ additional bad restrictions. Thus
every $(l,L,r_{n-2},\cdots ,r_1)$
reduces to Theorems \ref{thm1} and \ref{thm2}.
In a similar way, one can show that for other restrictions of type
$(l,L,r'_{n-2}, \cdots , r'_1 )$ we have $m^2 -m +1$ planes
while we need $m^2 -m$ planes.

{\it 2nd step. } Set $(l,L)=(n-2, (n-2)(m-1))$.
Such a restriction can be obtained
by dividing one of chains
\begin{equation}
z^{(k)}=z^{(k+1)}\tau^2=
\cdots =z^{(k+j)}\tau^{2j},
\label{eqn:old}
\end{equation}
of a restriction of type
$(l,L+1)$ into two pieces like
\begin{equation}
z^{(k)}=\cdots =z^{(k+i)}\tau^{2i}, {~~~\rm and ~~~}
z^{(k+i+1)}\tau^{2(i+1)}=\cdots =z^{(k+j)}\tau^{2j},
\label{eqn:new}
\end{equation}
where $0 \leq i \leq j-1$.

Note that there are two kinds of restriction planes for $(l,L)$,
{\it old ones} and {\it new ones} :
old ones are restriction planes which decrease
the degree of (\ref{eqn:e=1}) even for (\ref{eqn:old});
and new ones are those which does not decrease
the degree of (\ref{eqn:e=1}) for (\ref{eqn:old}) but
does decrease it for (\ref{eqn:new}).

For example, the following restriction of type $(l,L)$
\begin{equation}
\begin{array}{l}
z_1^{(3)}=z_1^{(4)}\tau^2
= \cdots =z_1^{(n)}\tau^{2(n-3)}=z_0 \tau^{2(n-2)}, \\
z_{i}^{(2)}=z_i^{(3)}\tau^2 =
\cdots = z_i^{(n)}\tau^{2(n-2)}, ~~ 2 \leq i \leq m-1, \\
z_m^{(2)}=z_m^{(3)}\tau^2 =
\cdots = z_m^{(n-1)}\tau^{2(n-3)}, \\
z_1^{(1)}=z_1^{(2)}\tau^2 ,
\end{array}
\label{eqn:exnew}
\end{equation}
can be obtained from the following restriction of type $(l,L+1)$
\begin{equation}
\begin{array}{l}
z_1^{(3)}=z_1^{(4)}\tau^2
= \cdots =z_1^{(n)}\tau^{2(n-3)}=z_0 \tau^{2(n-2)}, \\
z_i^{(2)}=z_i^{(3)}\tau^2 =
\cdots = z_i^{(n)}\tau^{2(n-2)}, ~~ 2 \leq i \leq m, \\
z_1^{(1)}=z_1^{(2)}\tau^2 ,
\end{array}
\label{eqn:exold}
\end{equation}
by resetting the relation between $z_m^{(n-1)}$ and $z_m^{(n)}$.
For (\ref{eqn:exold}) we have $m^2 -m+1$ restriction planes
\begin{equation}
\begin{array}{l}
z_j^{(2)}=z_1^{(3)}\tau^2 , ~~ 1 \leq j \leq m, \\
z_k^{(1)}=z_1^{(2)}\tau^2 , ~~ 2 \leq i, \leq m, 1 \leq k \leq m-1,
\end{array}
\label{eqn:exrest}
\end{equation}
which reduce to Theorems \ref{thm1} and \ref{thm2}.
For (\ref{eqn:exnew}) $m^2 -m+1$ restriction planes
(\ref{eqn:exrest}) are old ones.
On the other hand, restriction planes
$z_m^{(n-1)}=z_m^{(n)}\tau^2 $,
$z_1^{(2)}=z_m^{(n)}\tau^2 $ and
$z_k^{(1)}=z_m^{(n)}\tau^2 (1 \leq k \leq m-1)$ are new ones.

It is evident that the number of restriction planes
for $(l,L+1)$ and that of old restriction planes for $(l,L)$
coincide. Thus in order to show this case,
we need $m^2 -(m^2 -m+1)=m-1$ new restriction planes.

Let us consider the case $i=0$ in (\ref{eqn:new}).
Now there are $p$ free $z$'s belonging to
$\zeta^{(k)}$, and
there are $m-1-p$ free $z$'s belonging
to $\zeta^{(i)}$, where $i\neq k$,
and $1 \leq p \leq m-1$. Thus
we have at least $p+p(m-1-p) \geq m-1$ new restriction planes.
One can show the other case $1 \leq i \leq j-1$ similarly.
Hence this reduces to larger restrictions.

{\it 3rd step. } Suppose that we have enough number of restriction
planes to show (\ref{eqn:afterL}) for a restriction
of type $(n-2,L,r_{n-2}, \cdots , r_1 )$, where $L\leq (n-2)(m-1)$.
If we replace $L$ by $L-1$, we need $m$ new restriction planes.
On the other hand, there are now at least $m$ free $z$'s.
Thus, by a parallel argument
given in {\it 2nd step} this case reduces to
larger restriction.

{\it 4th step. } Suppose that we have enough number of restriction
planes to show (\ref{eqn:afterL}) for the restriction
of type $(l,L,r_{n-2},\cdots ,r_1)$,
where $L \leq (n-2)(m-1)$.
If we replace $l$ by $l-1$, we need $m$ new restriction
planes. Now we have at least $m$ free $z$'s.
Thus, from the same argument given in {\it 2nd step},
this case reduces to larger restrictions.
As for the restriction obtained by replacing $L$ by $L-1$,
the argument again perfectly parallels the one given in {\it 2nd step}.

Therefore, after repeating this procedure,
we reach
$$
(l,L,r_{n-2},\cdots ,r_1)= (0, \cdots , 0),
$$
and this is what we wish to prove.
$~~~\Box$

Next we notice that Proposition \ref{e=1} implies (\ref{eqn:cyclic})
for $\varepsilon_1 =i$.
For that purpose let us seek the residue formula for
$\overline{G}$.

~
Let
$$
G(z^{(n)}|\cdots |z^{(1)})=
\oint_{C} dx_1 \cdots \oint_{C} dx_M
\overline{G}(x | z^{(n)}|\cdots |z^{(1)})
\Psi (x | z),
$$
and
$$
\overline{G}_i (z^{(n)}|\cdots |z^{(1)})=
G^{\overbrace{n\cdots n}^m \cdots \hat{i}\cdots \hat{i}\cdots
   \overbrace{1\cdots 1}^m \overbrace{i\cdots i}^m}(z^{(n)}|\cdots |z^{(1)}).
$$
Then we have
\begin{equation}
\begin{array}{rcl}
\overline{G}^{(m)}_i (x|z) & = & 
(-\tau)^{i-1}\displaystyle\sum_{
1\leq \mu_1<\cdots <\mu_{n-1}\leq M}
        (-1)^{\sum_{i=1}^{n} mn-m-n+i +\mu_i } \\
& \times & 
\det\left( h_{\alpha}^{(m-1)}(x_{\mu_i}|z')\right) _{
1\leq i, \alpha \leq n-1}
\overline{G}^{(m-1)}_i (\stackrel{\mu}{\hat{x}}| z).
\end{array}
\end{equation}
Therefore (\ref{i=1, n}) for $i\neq 1$ reduces to Proposition \ref{e=1}.

Thus we have proved the Theorem \ref{mainthm}.

~

\section{Conclusions and discussions}

In this paper we present an integral formula
for quantum KZ equation of level $0$ associated with the
vector representation of $U_q ( \widehat{\frak s \frak l_n} )$.
This work is a generalization of our previous paper \cite{JKMQ}.
Smirnov obtained the formula for form factors
of the $SU(n)$ chiral Gross-Neveu model in Appendix A
in his book \cite{Smbk}.
It gives a rational scaling limit of the present work 
(i.e., $q=e^{\epsilon},~
z=e^{-\epsilon n \frac{\beta}{\pi i}}, \epsilon \to 0 $). 
Smirnov \cite{Smbk} has studied form factors of
integrable massive theories.
Equations (\ref{eqn:R-symm}) and (\ref{eqn:cyc})
are two of the axioms proposed by him
that form factors obey.
Let $G(z)$ be a form factor of a certain operator.
Then
a $\vartheta$
function in the integral kernel should be determined.
It is interesting and important to solve this problem.

Finally, we discuss the related works
on the integral formulas for
the quantum KZ equations.
In \cite{CORR}
an integral formula for correlation functions of the $XXZ$ model
was obtained on the basis of the
bosonization of the level $1$ highest weight representations
of $U_q ( \widehat{\frak s \frak l_2} )$.
This scheme gives only one particular solution
though those correlations satisfy the quantum KZ equation
of arbitrarily level.
A formula for higher spin analog of the $XXZ$ model
in terms of Jackson-type integral was given in \cite{KQS},
by using the level $k$ bosonization
of $U_q ( \widehat{\frak s \frak l_2} )$.

In \cite{Reshet,Matsuo} solutions by
Jackson-type integrals are obtained.
Their formulae are in principle valid for general level,
as opposed to our integral formula restricted to level $0$.
On the other hand, the problem of choosing the cycles for Jackson-type
integrals, which accommodates the freedom of the solutions,
is not well studied.

Tarasov and Varchenko \cite{TV} improved the Jackson-type integral formula
for $U_q ( \frak g \frak l_n )$ such that
the solutions automatically satisfy the $R$-matrix symmetry.
The number of Jackson-type integrals is $n(n-1)m/2$ for $N=nm$, whereas
our formula is written by $(n-1)m-1$ fold integral.
For $n=2$ two numbers are different by only $1$, however,
the difference gets greater as $n$ increases.

~

\section*{Acknowledgments}
The authors would like to thank M. Jimbo, T. Miwa and F. A. Smirnov
for a number of discussion.
Y.-H.Q. is partly supported by the Grant-in-Aid for
Scientific Research from the Ministry of Education,
Science and Culture, Japan,
No. 04-2297;
and by Soryushi Shogakukai.

~


\begin{thebibliography}{99}
\bibitem{JKMQ}Jimbo, M., Kojima, T., Miwa, T. and Quano, Y.-H.,
Smirnov's integrals and quantum Knizhnik--Zamolodchikov
equation of level $0$, J. Phys. {\bf A27} (1994) 3267--3283. 
\bibitem{FR}Frenkel, I. B. and Reshetikhin, N. Yu.,
Quantum affine algebras and holonomic difference equations,
Comm. Math. Phys. {\bf 146} (1992) 1-60.
\bibitem{Smbk}Smirnov, F. A.,
{\it Form Factors in Completely Integrable Models
of Quantum Field Theory},
World Scientific, Singapore, 1992.
\bibitem{Sm1}Smirnov, F. A.,
Dynamical symmetries of massive integrable models 1,2:
Int. J. Mod. Phys. {\bf 7A, Suppl. 1B} (1992) 813-837, 839-858.
\bibitem{CORR}Jimbo, M., Miki, K., Miwa, T. and
Nakayashiki, A., Correlation functions of the $XXZ$
model for $\Delta <-1$, Phys. Lett. {\bf 168A} (1992) 256-263.
\bibitem{KQS}Kato, A., Quano, Y.-H. and Shiraishi, J.,
Free boson representation for $q$-vertex operators
and their correlation functions,
Comm. Math. Phys. {\bf 157} (1993) 119-137.
\bibitem{Reshet}Reshetikhin, N. Yu.,
Jackson-type integrals, Bethe vectors, and
solution to a difference analog of
Knizhnik--Zamolodchikov system,
Lett. Math. Phys. {\bf 26} (1992) 153-165.
\bibitem{Matsuo}Matsuo, A., Quantum algebra
structure of certain Jackson integrals,
Comm. Math. Phys. {\bf 157} (1993) 479-498.
\bibitem{TV}Tarasov, V. and Varchenko, A.,
Jackson integral representations for solutions of the quantized
Knizhnik--Zamolodchikov equation, RIMS preprint, RIMS-949.
\end{thebibliography}
\end{document}